\documentclass[]{article}
\usepackage{lmodern}
\usepackage{amssymb,amsmath}
\usepackage{ifxetex,ifluatex}
\usepackage{fixltx2e} 
\ifnum 0\ifxetex 1\fi\ifluatex 1\fi=0 
  \usepackage[T1]{fontenc}
  \usepackage[utf8]{inputenc}
\else 
  \ifxetex
    \usepackage{mathspec}
  \else
    \usepackage{fontspec}
  \fi
  \defaultfontfeatures{Ligatures=TeX,Scale=MatchLowercase}
\fi
\IfFileExists{upquote.sty}{\usepackage{upquote}}{}
\IfFileExists{microtype.sty}{%
\usepackage{microtype}
\UseMicrotypeSet[protrusion]{basicmath} 
}{}
\usepackage[margin=1in]{geometry}
\usepackage{hyperref}
\hypersetup{unicode=true,
            pdftitle={Smart contracts for container based video conferencing services: Architecture and implementation},
            pdfauthor={Sandi Gec, Dejan Lavbič, Marko Bajec and Vlado Stankovski},
            pdfborder={0 0 0},
            breaklinks=true}
\usepackage{natbib}
\usepackage{longtable,booktabs}
\usepackage{graphicx,grffile}
\makeatletter
\def\maxwidth{\ifdim\Gin@nat@width>\linewidth\linewidth\else\Gin@nat@width\fi}
\def\maxheight{\ifdim\Gin@nat@height>\textheight\textheight\else\Gin@nat@height\fi}
\makeatother
\setkeys{Gin}{width=\maxwidth,height=\maxheight,keepaspectratio}
\IfFileExists{parskip.sty}{%
\usepackage{parskip}
}{
\setlength{\parindent}{0pt}
\setlength{\parskip}{6pt plus 2pt minus 1pt}
}
\providecommand{\tightlist}{%
  \setlength{\itemsep}{0pt}\setlength{\parskip}{0pt}}
\ifx\paragraph\undefined\else
\let\oldparagraph\paragraph
\renewcommand{\paragraph}[1]{\oldparagraph{#1}\mbox{}}
\fi
\ifx\subparagraph\undefined\else
\let\oldsubparagraph\subparagraph
\renewcommand{\subparagraph}[1]{\oldsubparagraph{#1}\mbox{}}
\fi

\let\rmarkdownfootnote\footnote%
\def\footnote{\protect\rmarkdownfootnote}

\usepackage{titling}

  \title{Smart contracts for container based video conferencing services:
Architecture and implementation}
    \pretitle{\vspace{\droptitle}\centering\huge}
  \posttitle{\par}
    \author{Sandi Gec, Dejan Lavbič, Marko Bajec and Vlado Stankovski}
    \preauthor{\centering\large\emph}
  \postauthor{\par}
    \date{}
    \predate{}\postdate{}

\usepackage{booktabs}
\usepackage{longtable}
\usepackage{array}
\usepackage{multirow}
\usepackage{wrapfig}
\usepackage{float}
\usepackage{colortbl}
\usepackage{pdflscape}
\usepackage{tabu}
\usepackage{threeparttable}
\usepackage{threeparttablex}
\usepackage[normalem]{ulem}
\usepackage{makecell}
\usepackage{xcolor}

\usepackage{amsthm}

\theoremstyle{definition}

\theoremstyle{definition}

\theoremstyle{definition}

\theoremstyle{remark}

\let\BeginKnitrBlock\begin \let\EndKnitrBlock\end
\begin{document}
\maketitle

\begin{quote}
Sandi Gec, \textbf{Dejan Lavbič}, Marko Bajec and Vlado Stankovski.
2018. \textbf{Smart contracts for container based video conferencing
services: Architecture and implementation},
\href{http://2018.gecon-conference.org/}{15th International Conference
on the Economics of Grids, Clouds, Systems, and Services \textbf{(GECON
2018)}}, 18. 9. 2018 - 20. 9. 2018, Pisa, Italy.
\end{quote}

\section*{Abstract}\label{abstract}
\addcontentsline{toc}{section}{Abstract}

Today, container-based virtualization is very popular due to the
lightweight nature of containers and the ability to use them flexibly in
various heterogeneously composed systems. This makes it possible to
collaboratively develop services by sharing various types of resources,
such as infrastructures, software and digitalized content. In this work,
our home made video-conferencing (VC) system is used to study resource
usage optimisation in business context. An application like this, does
not provide monetization possibilities to all involved stakeholders
including end users, cloud providers, software engineers and similar.
Blockchain related technologies, such as Smart Contracts (SC) offer a
possibility to address some of these needs. We introduce a novel
architecture for monetization of added-value according to preferences of
the stakeholders that participate in joint software service offers. The
developed architecture facilitates use case scenarios of service and
resource offers according to fixed and dynamic pricing schemes, fixed
usage period, prepaid quota for flexible usage, division of income,
consensual decisions among collaborative service providers, and
constrained based usage of resources or services. Our container-based VC
service, which is based on the Jitsi Meet Open Source software is used
to demonstrate the proposed architecture and the benefits of the
investigated use cases.

\section*{Keywords}\label{keywords}
\addcontentsline{toc}{section}{Keywords}

Blockchain, Video-Conferencing, Container, Monetization, Smart Contracts

\section{Introduction}\label{introduction}

Resources, such as computing infrastructures, Cloud services offers and
software (Web servers, libraries and so on), combined together represent
basis for the provisioning of high-quality software services. The
various stakeholders usually need to collaborate in order to be able to
produce such software services. This particularly requires mechanisms
for assuring monetization of the contributed added-value, transparency,
security, trust, Quality of Service (QoS) assurance through Service
Level Agreements (SLAs), and so on.

In order to facilitate flexible provisioning and consumption of
resources and services, it is necessary to address various SLAs and
monetization use cases of the stakeholders. In this study we concentrate
on the analysis and implementation of various monetization approaches to
collaboratively provide comprehensive software services to the end
users. This includes SLA offers according to a fixed and variable
pricing model, time-limited and quota-based provisioning,
constraints-based access, revenue sharing, and other mechanisms making
it possible to engineer and deliver software services with sufficient
business flexibility. The goal of the present study is therefore to
develop and evaluate an architecture that can be used to automate the
process of software services provisioning and consumption.

Our approach relies on recent trends in the financial domain.
Traditional currency transactions among people and companies are often
facilitated with a central entity and controlled by a third party
organizations such as banks. Bitcoin as a first decentralized digital
currency was presented and launched as an alternative to centralized
solutions. It is based on the Blockchain technology and has many
benefits \citep{anjum_blockchain_2017}. Through the adoption of Bitcoin
among the world population the Blockchain technology presents
opportunities in many areas, such as the Internet of Things, Cloud
computing and software engineering in general. This has led to the
launch of new dedicated cryptocurrencies -- Ethereum\footnote{\url{https://www.ethereum.org/}}
as a Smart Contracts (SC) ledger, IoTa\footnote{\url{https://iota.org/}}
as an Internet of Things-based (IoT) ledger, Ripple\footnote{\url{https://ripple.com/}}
as a transaction cost optimization ledger, ledgers for anonymous
transactions, protocols and other Blockchain systems covering a plethora
of use cases. For example, in the Cloud domain the distributed storage
system StorJ\footnote{\url{https://storj.io/}} is an alternative to
commercial storage solutions such as Amazon S3 storage, Google Drive or
Dropbox, and provides the same types of services with high encryption
security and full transparency.

In this paper, we propose a new Cloud architecture for Container Image
(CI) management containing a Web based Video-Conferencing (VC)
application that assures high QoS and exploits the benefits of the
Blockchain technology as a key component that differs from other
existing solutions. Some key technologies, such as SCs, that are part of
the architecture, are used to fulfill functional requirements, such as
the needed agreements among the user and system roles. These are
described in the following sections. By developing solutions for several
generic monetization use cases, we aim to cover a potentially large
number of stakeholders accordingly with their usage preferences, such as
regular, occasional, demanding users and similar. This new architecture
is designed having in mind the requirements for security, transparency
and various monetization approaches through the use of SCs. The new
architecture is implemented for one VC applications on top of
Blockchain. This makes it possible to empirically compare it with
traditional monetization approaches quantitatively and qualitatively.

The rest of the paper is organized as follows. Section
\ref{related-work} positions our work among other related works. Section
\ref{economic-requirements} introduces Blockchain and Cloud computing in
relation to the economy aspect. Section \ref{blockchain-use-cases}
presents and overview of the use cases based on using Blockchain.
Section \ref{blockchain-based-architecture} presents the new
architecture and its implementation. Section \ref{in-depth-analysis}
presents a detailed overview of dynamic price monetization and explains
the detailed workflow among the software services. Section
\ref{discussion-conclusion} explains the relevance and significance of
the obtained results, discusses the lessons learnt and provides some
conclusions.

\section{Related work}\label{related-work}

\citet{haber_how_1991} presented a theoretical idea of the Blockchain
concept, on how to certify digital documents in order to assure the
tamper-proof data integrity. The first practical attempt of Blockchain
technology was presented with the launch of the cryptocurrency Bitcoin
\citep{nakamoto_bitcoin:_2008} in 2009. Due to the Bitcoin simplicity of
just sending and receiving digital assets, many researchers and
Blockchain enthusiasts launched their own Blockchain cryptocurrencies.
An interesting concept was presented by Vitalik et al.
\citep{buterin_ethereum_2015} with the introduction of the Turing
complete Smart Contracts (SCs) that can be compared to general (notary)
contracts with limited, but at the same time sufficient functionalities
that may cover several different use cases.

Blockchain monetization potentials were presented through practical
applications and latest global trends by \citet{swan_anticipating_2017}.
The most potential applicable areas are digital assets registry
management, solving the issue of billions of ``unbanked'' people,
long-tail personalized economic services and payment channels in terms
of creation of financial contracts executed over time.
\citet{yoo_blockchain_2017} analysed the potential of the Blockchain
technology from the global financial perspective and thus also leaning
on the use case of micro payments. A comparison between the current
monetary system and Blockchain based is objectively presented by
\citet{ankenbrand_structure_2018}. On the other hand
\citet{peter_blockchain-applications_2017} outline the actual
operational and regulatory challenges in terms of scalability,
interoperability, standards, governance and others which should be
defined with governments and financial institutes. All of these works
explore ideas from a high-level perspective and do not provide any
concrete monetization use cases, which are presented in our work.

Cloud computing has been dramatically adopted in all information
technology (IT) environments for its efficiency, availability and
hardware resource scalability. Therefore, Cloud computing architectures
usually consist of variety of sub-systems, such as front-end and
back-end platforms, Cloud based delivery and networks that are designed
to address specific end user needs and requirements. Various
requirements are also directed towards achieving high Quality of Service
(QoS) and Quality of Experience (QoE). Some studies focus on the design
of Cloud based orchestration systems that provide an intelligent
delivery of CI based applications, e.g.~a File Upload and a VC
application as presented ih our previous work
\citep{pascinski_qos-aware_2018}. This, however, is a first study that
investigateds the needs for monetization in the Cloud services economy
which relies on the use of Blockchain. Our study aims to increase the
overall system robustness in terms of system distribution, when various
software services are developed, engineered, deployed and operated.
Several highly focused studies have used Blockchain technology to
improve product traceability and quality preservation, such as the study
of \citet{lu_adaptable_2017}. The work of \citet{xia_medshare:_2017}
focused on adoption of Blockchain for trust-less medical data sharing as
a State-of-the-Art solution in the domain of medical Big Data. On the
other hand the overall performance of such systems has not been
addressed sufficiently and they have not been described in the context
of SCs. An attempt of integration of SCs in an existing Cloud system for
VMI and CI management was presented on H2020 ENTICE\footnote{\url{http://www.entice-project.eu/}}
project \citep{gec_semantics_2018} as an agreement management component
among users. The present work builds on top of the benefits of SCs, in
order to define, develop and test various monetization strategies which
may be useful in the domain of the Cloud services economy.

\section{Background}\label{economic-requirements}

Blockchain and SCs are technologies underpinning the design and
implementation of our new architecture on top of which our Jitsi
Meet\footnote{\url{https://jitsi.org/}} based VC system is provisioned.
Blockchain (originally named block chain), is a continuously
(time-based) growing list of records, called blocks, which are linked
and secured using cryptography mechanisms (e.g.~Bitcoin uses SHA-256).
Each record ordinarily contains metadata -- a reference (cryptographic
hash) to the previous block, timestamp and transaction data. Blockchains
are secure \textbf{by design} through broker-free (P2P-based)
characteristics. Therefore, the alteration of the content in the blocks
by single or even multiple node entities is very difficult to happen due
to the high distribution rate. To summarize, key properties of the
Blockchain are elevated distribution, no central authority,
irreversibility, accessibility, time-stamping and cryptography.

Smart Contract (SC) is a digital variation of a traditional contract
which can be described as a protocol intended to digitally facilitate,
verify and enforce the negotiation or performance by implementing
arbitrary rules. In the design of our novel architecture, we rely on the
Ethereum network with the built-in fully fledged Turing-complete
programming language Solidity, which is used to design and implement
SCs.

\begin{figure}

{\centering \includegraphics[width=1\linewidth]{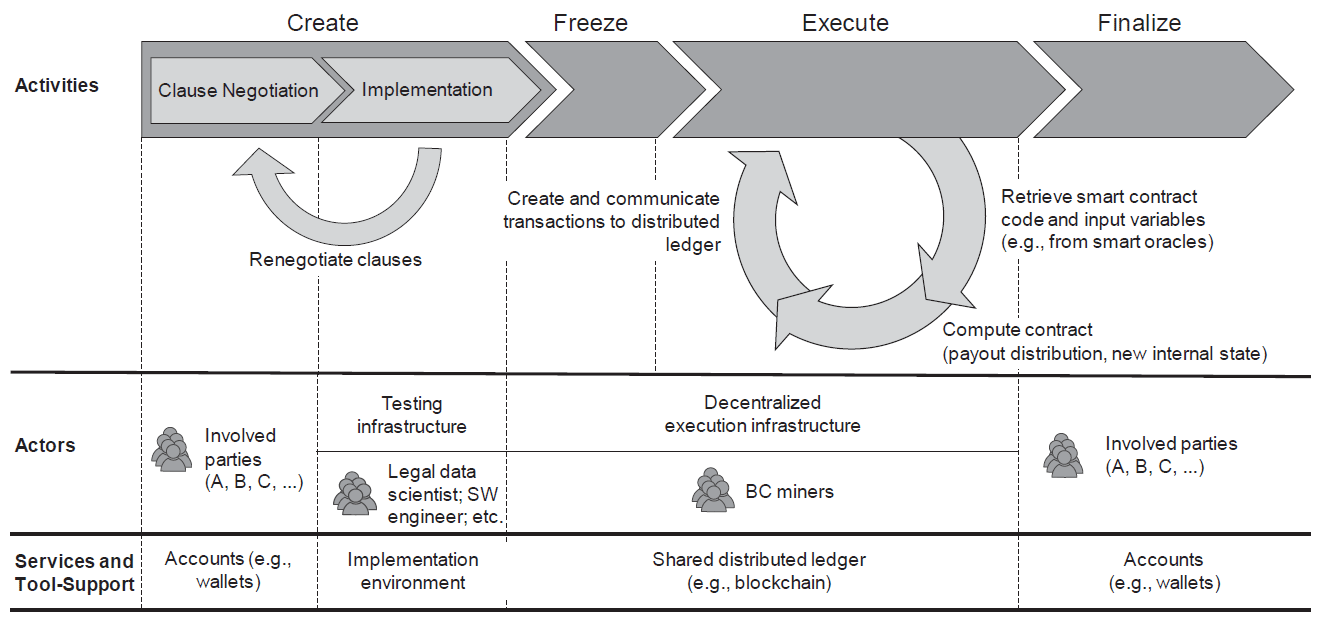}

}

\caption{The life cycle of a smart contract: phases, actors, and
services \citep{sillaber_life_2017}}\label{fig:life-cycle}
\end{figure}

The life-cycle of a SC can be summarized in the following example. A
developer designs a dedicated SC or uses an existing one. By doing that
SC is considered as a template that is desired to be at least validated
through specific tools. Each SC template may be deployed on a desired
address to a production (mainnet) or development (testnet) network. Upon
deployment the SC invokes the constructor only once, while the address
owner usually has higher privileges, for example, it can destroy the
deployed SC instance. Other participant, public or specific addresses,
can trigger a SC instance through the supported functions that lead to
sending of another transaction, triggering another SC, theoretically
\textbf{ad infinitum} as illustrated in Figure \ref{fig:life-cycle}.

In order to understand the monetization in the Cloud services domain we
first need to know the economic ecosystem of the resources, services and
environments deployed in the Clouds. In the first place, such
environments must consider the total cost of ownership for the
on-premises by considering the cost of the resources, equipment,
computing and networking infrastructures, capital and the resources and
services lifespan. All of these aspects can be described as the
operational and maintenance costs of the services running on demand.
Currently, there exist various special-purpose monetization services
available, such as YouTube channel monetization, mobile applications
advertising monetization and similar, which commonly have relatively
expensive operational costs. It may therefore be possible to implement a
flexible Blockchain monetization overlay on top of Cloud services.

In this paper, we focus on several use cases and requirements for
monetization of a specific container-based VC service as an
architectural overlay that may satisfy different types of end users. The
developed architecture and its implementation is described in detail in
the following section \ref{blockchain-use-cases}.

\section{Video-conferencing application, use cases and requirements
analysis}\label{blockchain-use-cases}

In this work we rely on advanced technologies, which are used in Cloud
computing including Docker\footnote{\url{https://www.docker.com/}} for
the management of containers, and Kubernetes\footnote{\url{https://kubernetes.io/}}
as a general purpose orchestration technology. Our VC application Jitsi
Meet is therefore packed into Docker containers. In contrast to VMIs,
containers can be spinned on and off much faster, practically within
seconds, thus forming basis for fine-grained services orchestration,
which is driven by various events (such as end users that need to run a
video-conference). Whenever a specific software service is required,
these services make it possible to dynamically deploy a container image
and serve the particular end user or event.

Our goal was to study different monetization approaches that can be used
by the stakeholders that contribute resources and services in order to
develop a working VC software service. We present a basic comparison
between conventional and Blockchain monetization on the Ethereum
network.

In order to ensure feasibility of the monetization method applied in our
system, we summarized basic properties of the Ethereum network and
compared them with three popular payment systems: Visa, Mastercard and
Paypal. The comparison results in Table
\ref{tab:different-monetization-methods} show that the overall cost of
Ethereum monetization is significantly lower compared to traditional
methods. Ethereum transaction cost represented with GWEI unit where
\emph{\(1\) ETH = \(10^9\) GWEI}, is tightly dependent by the following
factors: Ethereum network congestion, preferred transaction speed and
the actual price of ETH coin. To minimize the ETH coin volatility there
is a dedicated ERC20 token TrueUSD\footnote{\url{https://www.trusttoken.com/}},
which is an USD-backed, fully collateraized, transparent and legally
protected. In addition to lower absolute transaction cost, SCs allows
much more flexible transactions. For example, lock-in of transaction
funds refers to the actual locking of funds in the SC till certain
functional conditions has not been reached and releasing the funds, or
just fractioned funds, to the appropriate user entity.

\begin{table}[t]

\caption{\label{tab:different-monetization-methods}Main properties of different monetization methods in the first half of the year 2018}
\centering
\begin{tabu} to \linewidth {>{\raggedright\arraybackslash}p{3cm}>{\raggedright\arraybackslash}p{3cm}>{\raggedright\arraybackslash}p{4cm}>{\raggedright\arraybackslash}p{3cm}}
\toprule
Monetization method & Transaction processing fee & Merchant or operational service cost [USD] & Lock-in of transaction funds\\
\midrule
Visa & \(1.43\%\) -- \(2.40\%\) & min. \(1.25\%\) & limited\\
Mastercard & \(1.55\%\) -- \(2.60\%\) & min. \(1.25\% + 0.05\) & limited\\
PayPal & \(2.90\%\) -- \(4.40\%\) & min. \(1.50\%\) & limited\\
Ethereum & \(1\) -- \(40\) GWEI\footnote{\url{https://kb.myetherwallet.com/gas/what-is-gas-ethereum.html}} & \(0\) & flexible\\
\bottomrule
\end{tabu}
\end{table}

The monetization processes for our VC system from the viewpoint of the
various stakeholders was described with several use cases as follows.
The use cases describe different user needs based on actual usage
patterns of resources and services. These can be achieved through the
definitions of SCs. Another important aspect, which is considered
concerns the income division among the parties, including the Cloud
providers, container image deployment platforms, infrastructure
providers and so on. In order words, we analyed the needs of all
stakeholders who are in charge for an individual VC service setup
processes. Following are seven identified monetization use cases that
can be supported by Blockchain and Ethereum SCs.

\textbf{Fixed price} is the most basic definition of the use case. The
VC system, depending on the QoS end user's requirements, offers the
usage of VC service for a fixed price and fixed maximum period of time
on demand. The agreement is reached when both parties signs the SC while
the end user sends the required ETH funds.

\textbf{Dynamic price} in addition to fixed price use case, offers
higher flexibility of the actual VC service availability. For example,
the end user knows the maximum time that that he/she might need the VC
service. After the agreement is reached the user pays the full price but
the Ethereum funds are locked by the SC. The fund become unlocked if the
maximum period is reached or if the end user stops using VC service by
confirming the SC. In the unlocking phase the actual usage time is
charged -- the proportional Ethereum of unused time is returned to the
end user while the rest is sent to the VC service.

\textbf{Time-limited usage} refers to the use case when the end user in
advance agrees the per-minute conversation price, buys certain VC
session minutes and uses the VC service gradually on demand. A typical
usage is suitable for Big Brother-like TV shows and other content
provided in real-time online.

\textbf{Flexible usage period} offers more flexibility for the end users
that cannot define the exact period when the VC service will be used. In
the proposed end user's period, the VC service has to be either
available or the deployment of the VC service has to be optimized and
thus very fast. The charge methodology follows the dynamic price use
case and the SC contains also the minimum charge price for maintaining
the fast deployment of the containerized VC service for the specific
time period.

\textbf{Division of income} is a particular monetization process where
the parties involved are those who enables the VC service hardware and
software infrastructure. For example, one provider offers specific VC
containers, another offers infrastructure and the third provider offers
other services, such as monitoring. The division of income is agreed
upon the parties in advance and validated through the SC. An overall
advantage of this approach for the end user is reflected as a lower
overall price of the VC service.

\textbf{Consensus decision} is an upgrade to division of income use
case. The consensus is reached through a democratic voting SC, similar
to the one proposed by \citet{bragagnolo_smartinspect:_2018}. As a
result the management of VC service is divided among parties specialized
for different HW/SW and service aspects and thus improved QoS/QoE,
availability and overall cost reduction.

\textbf{Constraint based} focuses on the legal aspects determined by the
end user's constraints. As an example EU General Data Protection
Regulation (GDPR) compliant can be agreed by all end users through SC.
In some cases geographical definition is also important and should be
determined by the end user and VC system in advance (e.g.~not all data
services are allowed in specific countries). In general, by constraint
limiting, either the price for the end user, either the potential QoS,
changes (VC service price increase, decreased QoS) in the proposition of
SC by the VC service.

In order to summarize, the rough analysis of the main properties of the
monetization use cases addressing the various requirements are presented
in Table \ref{tab:different-monetization-methods-use-cases}.

\begin{table}[t]

\caption{\label{tab:different-monetization-methods-use-cases}Main properties of different monetization use cases}
\centering
\begin{tabu} to \linewidth {>{\raggedright\arraybackslash}p{2cm}>{\centering\arraybackslash}p{2cm}>{\raggedright\arraybackslash}p{7cm}>{\raggedright\arraybackslash}p{3cm}}
\toprule
Monetization use case & Minimum required SCs & SCs functions [description (SC)] & Roles involved\\
\midrule
Fixed price & \(1\) & set price by owner, sign agreement & VC system, end user\\
Dynamic price & \(1\) & set price by owner, sign agreement, stop VC service & VC system, end user\\
Time-limited usage & \(1 + n\) (\(n\) are number of accesses) & set price by owner \((1)\), sign agreement
\((1)\), start VC session \((n)\), stop VC session \((n)\) & VC system, end user\\
Flexible period & \(1\) & set price by owner, sign agreement, stop VC service & VC system, end user\\
Division of income & \(2\) & set price by all VC enabling entities \((1)\),
set price by owner \((2)\), sign agreement \((2)\) & VC service enabling entities, end user\\
\addlinespace
Consensus decision & \(2\) & set voters \((1)\), vote by all VC enabling
entities \((1)\), set price by owner \((2)\), sign agreement \((2)\) & VC service enabling entities, end user\\
Constraint based & \(1\)) & delegate constraints by owner, set price by owner, sign agreement, stop VC service & VC system, end user\\
\bottomrule
\end{tabu}
\end{table}

\section{Blockchain based architecture for flexible monetization of
Cloud resources and software
services}\label{blockchain-based-architecture}

The proposed architecture consist of various components that are packed
as Docker CI. In order to address the proposed monetization use cases,
we extend the existing VC architecture which includes an orchestrator
\citep{pascinski_qos-aware_2018}. While the focus of our previous work
was the design of a Cloud based orchestration system that provides an
intelligent delivery of two network intensive applications (File Upload
and VC), the focus of the present work addresses the needs to facilitate
flexible usage of the resources and software related to the provisioning
of the VC service. We present an architecture to facilitate the VC
application through the use of the Blockchain technology.

The novel architecture enables different monetization approaches among
four pillar component layers:

\begin{enumerate}
\def\labelenumi{\arabic{enumi}.}
\tightlist
\item
  Cloud providers,
\item
  Implemented solution services,
\item
  Blockchain components and
\item
  Graphical User Interface (GUI).
\end{enumerate}

\begin{figure}

{\centering \includegraphics[width=0.6\linewidth]{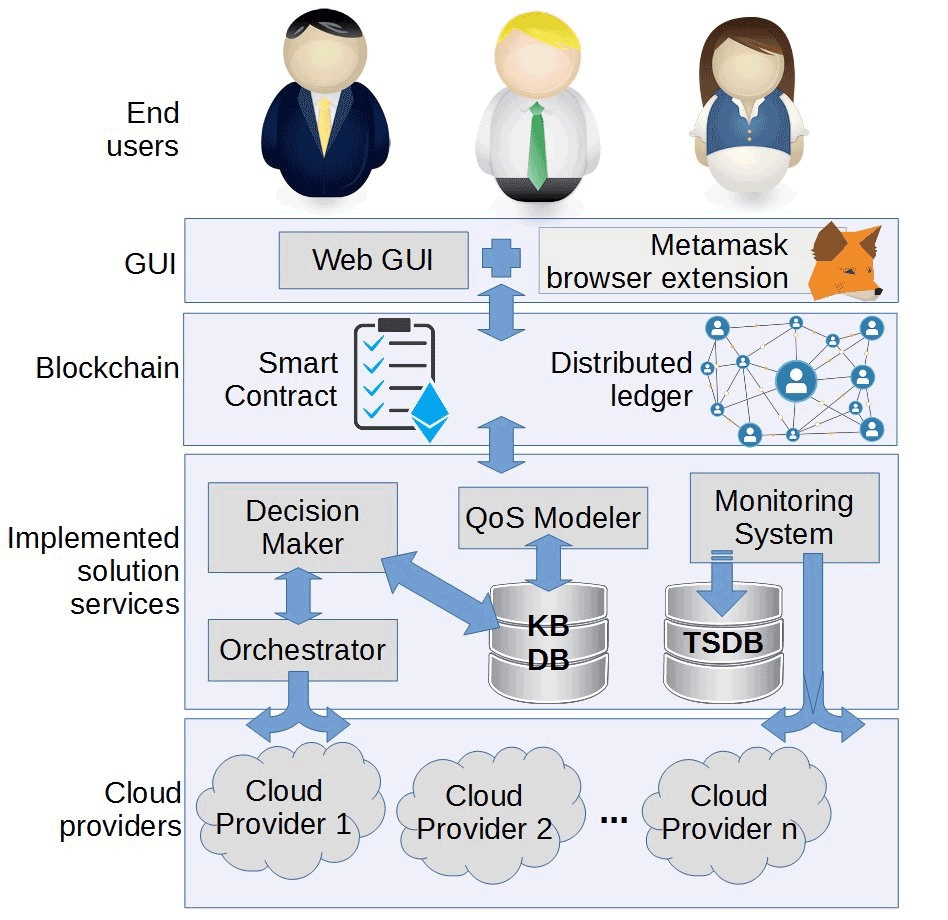}

}

\caption{High-level architecture of a SaaS VC use case}\label{fig:high-level-architecture}
\end{figure}

The implemented architecture is depicted in Figure
\ref{fig:high-level-architecture} and its main components are explained
in the following.

\textbf{Cloud providers} are available on different geographical
locations and are used for the deployment of VC service instances. The
VC application components, which are packed into Docker containers can
be deployed on demand when an agreement between the end user and the VC
service provider is reached, and thus follow the SaaS delivery method.
Besides, the deployed VC application instances can be monitored to
provide metrics, such as usage time period and other QoS metrics that
can be included in an SLA.

\textbf{Implemented service solutions} are composed of different
components that comprehensively perform different tasks. For example, to
estimate the potential QoS of the deployed VC service it is mandatory to
have QoS metrics defined and measured, while the VC service is running.
By doing this, the QoS metrics are modelled, while the Decision Maker
identifies potential Cloud providers, where the container image should
be started.

\textbf{Blockchain components} consist of a public Blockchain ledger,
i.e.~Ethereum and our developed SC templates supporting the proposed
monetization use cases. See Section \ref{blockchain-use-cases} for
details. The distributed node infrastructure is maintained by the
Ethereum community and offers a high level of distribution and
availability. The main API technologies are written in the Java
programming language, thus the core Ethereum bridge is the Java library
ethereumj\footnote{\url{https://github.com/ethereum/ethereumj}} embedded
in the core layer API. The deployed SC instances may be triggered by end
users and VC system. In all SC trigger executions and phases are updated
by the actual content in the Ethereum blockchain network.

\textbf{Web based GUI} is the main entry point for the end user. Each VC
session consists of an end user subscriber and other participants. The
subscriber in the Web GUI defines the preferences with QoS metrics and
submits the request that is delegated to the backend component.
According to the end user's preferences, the SC terms are proposed to
the end user by our system. In all monetization use cases the agreement
is reached when both parties signs the SC.

The general workflow of the architecture consist of the agreement among
an end user and the VC service to determine the monetization use case,
QoS user's requirements and the price. After the VC service deploys the
CI application components, the end user gets unique URL for the VC
session. The application life-cycle is completed when the end user
finishes using the VC application or if any other condition occurs,
e.g.~the VC session time is exceed. In addition the resources are freed
by the undeployment of VC session service on the Cloud providers'
infrastructure. In the final stage the Ethereum SCs are signed by both
parties, the end user and the VC service, and thus the final
monetization flow is executed. A complete workflow with the dynamic
price monetization use case is presented in the following section
\ref{in-depth-analysis}.

\section{In-depth analysis of the dynamic price monetization use
case}\label{in-depth-analysis}

In this section, we systematically describe the dynamic price
monetization workflow, which is used by our VC software service. The
selected use case is motivated by the transparent definition of the
agreements among end user and VC system through an Ethereum SC.

The aim of the dynamic price monetization is increased flexibility for
the end user, who cannot estimate the exact duration of the needed VC
session. In comparison to traditional monetization methods (e.g.~Paypal,
Mastercard, Visa etc.), which primarily perform as a trustful entity and
do lack transparency towards the end user, the SC Blockchain approach on
the other hand is transparent, without central authority, and uses the
defined functions in the Solidity programming language.

We integrated the SCs in our VC system in order to better understand the
dynamic price monetization. The entities involved in the process are:

\begin{enumerate}
\def\labelenumi{\arabic{enumi}.}
\tightlist
\item
  the end user who initiates the VC service,
\item
  the Blockchain component which provides SC execution,
\item
  the solution services, and
\item
  Cloud providers that actually provide the running infrastructure for
  the VC application components.
\end{enumerate}

\begin{figure}

{\centering \includegraphics[width=1\linewidth]{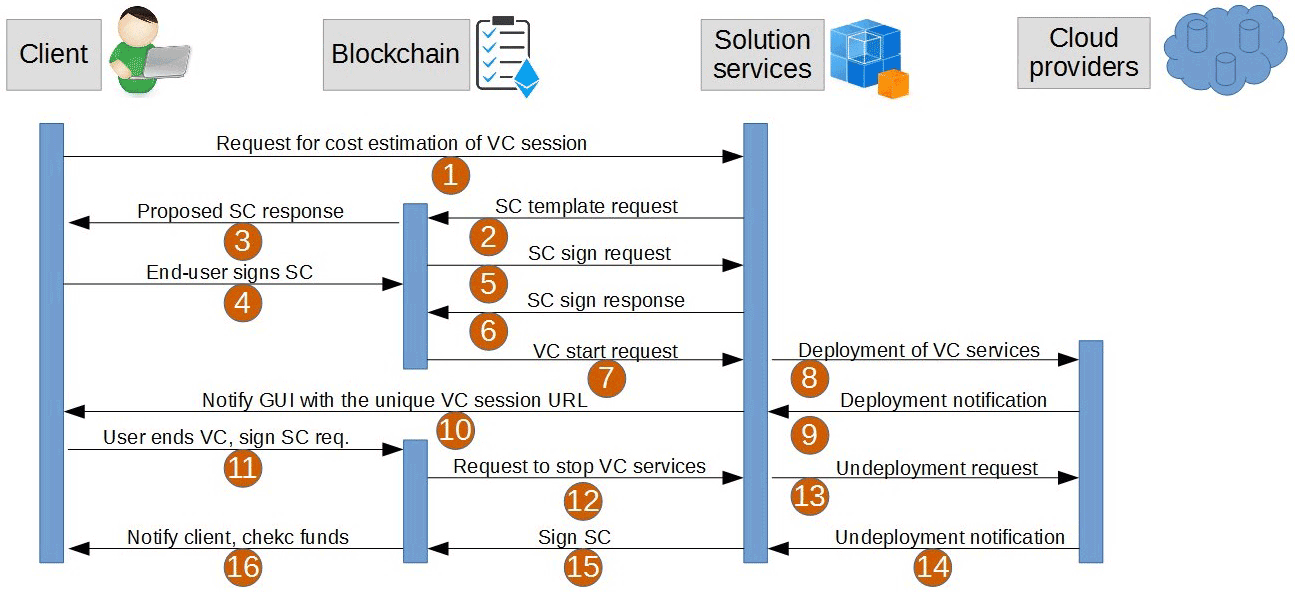}

}

\caption{Sequence diagram of the general VC use case}\label{fig:vc-sequence-diagram}
\end{figure}

The flow of interactions between the Blockchain component and the
triggering entities is depicted in Figure \ref{fig:vc-sequence-diagram}.
In the beginning the end user sets basic properties through the Web GUI,
for example the monetization type (dynamic price monetization), the
maximum time period (one hour) and determines additional QoS preferences
(availability \(99.8\%\), video quality high-definition).

The process flow is explained in the following.

\begin{enumerate}
\def\labelenumi{\arabic{enumi}.}
\tightlist
\item
  Based on the end user's properties, the VC solution services estimate
  the price for the end user, based on QoS/QoE history trends, and
  deploy an SC instance,
\item
  The notification about the pricing policy is passed to the end user,
\item
  who approves the SC agreement by paying the full incentive, locked in
  the SC, for one hour of Ethereum funds (ETH) through the Ethereum
  bridge MetaMask\footnote{\url{https://metamask.io/}}, used as Web
  browser extension and monetization interface for ETH cryptocurrency.
\item
  The signed SC notification is passed to the VC solution services,
\item
  where it is signed by VC service address, and
\item
  further passed as a deployment request to the solution services, to be
  more concrete the QoS Modeler and the Decision Maker.
\item
  Further on, the deployment process is executed, and
\item
  on success,
\item
  the end user gets an unique Uniform Resource Locator (URL),
\item
  that can be shared among other VC participants.
\item
  Although it is a very common practice to share unique public URLs,
  this can be additionally enhanced, for example by registering
  individual participants addresses to obtain the VC session access. The
  VC session ends by the SC signature of the end user or the session
  automatically stops if one hour period is exceeded and no signing of
  SC is needed
\item
  In the final stage the undeployment of CI application instances is
  executed.
\item
  On success the final SC function, which unlocks ETH funds and executes
  proportional (time based) return of the ETHs or no return in case when
  time limit is exceeded. This step could be skipped but it is
  recommended as a final security check in case of any VC service
  anomalies (e.g.~end user gets full refund in case when the
  availability of VC service dropped below \(75\%\)).
\item
  see 13
\item
  see 13
\item
  Finally, the end user gets the notification about the completion of
  the SC.
\end{enumerate}

In order to outline the SC template for the dynamic price monetization,
it is mandatory to define the key global attributes -- lock time in
seconds determined by current block timestamp, session start time,
maximum price (refereed to lock time in seconds), end user address and
VC owner address. Due to the usage of block based generation of the
timestamp, the overall delay is inducted between the discrete time
generation of blocks which is on average 15 seconds\footnote{\url{https://etherscan.io/chart/blocktime}}.

An important aspect is the management of event triggers which are not
supported in Ethereum SCs. A good practice recommendation is to either
develop the service on our own or simply use Ethereum's Alarm Clock
service\footnote{\url{http://www.ethereum-alarm-clock.com/}}, which is
designed for SC event purposes.

A fundamental SC template that locks the ETH funds of the sender is
described in Algorithm \ref{def:sc-lock}.

\begin{align*}
& \textbf{Data} \text{: object payable \{addressFrom, value\} } \\
& \textbf{Result} \text{: boolean success} \\
\\
& \textbf{if } basic\_conditions\_check \textbf{ then} \\
& \quad \text{// in case the ETH value does not match the agreed price} \\
& \quad \text{return false;} \\
& \textbf{else} \\
& \quad \text{// set the release time, address of the sender and deposit ETHs,} \\
& \quad \text{// while the lockTimeSeconds is a global attribute} \\
& \quad \text{releaseTime = currentBlockTime + lockTimeSeconds;} \\
& \quad \text{balance = balance + value;} \\
& \quad \text{globalAddressFrom = addressFrom;} \\
& \quad \text{return true;} \\
& \textbf{end}
\end{align*}

\BeginKnitrBlock{definition}
\protect\hypertarget{def:sc-lock}{}{\label{def:sc-lock} }SC function for
temporary locking of Ethereum funds.
\EndKnitrBlock{definition}

Another pillar function is the VC service stopping notification that is
triggered by the end user. After the successful trigger the ETH
transaction to the VC service address is performed and another one to
the address of the end user accordingly to the duration of the VC.

\section{Discussion and conclusion}\label{discussion-conclusion}

The presented monetization approach in the previous section is overall
less expensive compared to traditional monetization approaches. Beside
the lower and relative fee policy not varying on the amount of funds to
be sent, there are only initial integration costs and basically no
further operational costs. Deployed and active SCs that temporary store
the ETH funds of the end users, are accessible from the VC system APIs.
However, the address containing the processed ETH funds owned by the VC
system is stored in a cold storage wallet that make it secure in case of
potential attacks.

This study presents a novel architecture that encapsulates Cloud
principles and monetization possibilities through the usage of
Blockchain. It is shown that SCs are suitable for establishing
transparent mutual agreements among end users and VC system, and
therefore facilitate the overall payment process without any additional
cost for the service owners (e.g.~bank, payment cards and other fees).
Despite that, the specific SCs developed for the service needs to be
carefully designed and validated (e.g.~by using Oyente SC validation
tool\footnote{\url{https://github.com/melonproject/oyente}}), analysed
and evaluated at different levels, such as code pattern comparison and
simulation of actual usage with multiple parties on the same SC.

Following the implementation of our new architecture, we aim to
investigate the QoS of the described monetization capabilities through
performance measurements -- actual transaction costs, block confirmation
durations and other metrics related to the time metric.

The Cloud domain with Blockchain technology poses some very specific
research challenges for real-time applications (e.g.~VC) that need to
fulfil or enhance the functional requirements but at the same time
introduce new functionalities, such as monetization. In our use case
where the system methodology follows the pay-as-you-go concept, a
software is deployed on end user's demand and SCs are used as an
advanced agreement management tool among all stakeholders.

\section*{Acknowledgment}\label{acknowledgment}
\addcontentsline{toc}{section}{Acknowledgment}

This project has received funding from the European Union's Horizon 2020
Research and Innovation Programme under Grant Agreement No. 815141
(DeCenter project: Decentralised technologies for orchestrated
cloud-to-edge intelligence).

\end{document}